\documentclass[apj]{emulateapj}

\usepackage{epsfig}
\usepackage{color}
\usepackage{graphicx}

\newcommand{\chandra}{{\it Chandra}}
\newcommand{\rxte}{{\it RXTE}}
\newcommand{\xte}{{\it RXTE}}

\newcommand{\fermi}{{\it Fermi}}

\newcommand{\xmm}{{\it XMM-Newton}}

\newcommand{\nustar}{\textit{NuSTAR}}

\newcommand\etal{{et al.}}

\newcommand\psr{PSR~B1821$-$24}
\def\farcm{\hbox{$.\mkern-4mu^\prime$}}
\def\farcs{\hbox{$.\!\!^{\prime\prime}$}}

\begin{document}

\shorttitle{\nustar\ Observations of Three Energetic Millisecond Pulsars}
\shortauthors{Gotthelf \& Bogdanov}

\title{ \textit{NuSTAR} Hard X-ray Observations of the Energetic \\ Millisecond Pulsars PSR B1821$-$24, PSR B1937$+$21, and PSR J0218$+$4232}

\author{E.~V.~Gotthelf\altaffilmark{1} and S.~Bogdanov}
\affil{Columbia Astrophysics Laboratory, Columbia University, 550 West 120th Street, New York, NY 10027, USA}

\altaffiltext{1}{Departament de F\'{\i}sica Qu\`antica i Astrof\'{\i}sica, Institut de Ci\`encies del Cosmos, Universitat de Barcelona, IEEC-UB, Mart\'{\i}\ i Franqu\`es 1, E-08028, Barcelona, Spain}

\begin{abstract}
  We present \textit{Nuclear Spectroscopic Telescope Array}
  (\textit{NuSTAR}) hard X-ray timing and spectroscopy of the three
  exceptionally energetic rotation-powered millisecond pulsars
  PSRs~B1821$-$24, B1937$+$21, and J0218$+$4232. By correcting for
  frequency and phase drifts of the \textit{NuSTAR} onboard clock, we
  are able to recover the intrinsic hard X-ray pulse profiles of all
  three pulsars with a resolution down to $\leq 15~\mu$s.  The
  substantial reduction of background emission relative to previous
  broadband X-ray observations allows us to detect for the first time
  pulsed emission up to $\sim$50 keV, $\sim$20 keV, and $\sim$25 keV,
  for the three pulsars, respectively.  We conduct phase-resolved
  spectroscopy in the 0.5--79 keV range for all three objects,
  obtaining the best measurements yet of the broadband spectral shape
  and high-energy pulsed emission to date.  We find extensions of the
  same power-law continua seen at lower energies, with no conclusive evidence for
  a spectral turnover or break.  Extrapolation of the X-ray power-law
  spectrum to higher energies reveals that a turnover in the 100 keV
  to 100 MeV range is required to accommodate the high-energy
  $\gamma$-ray emission observed with \textit{Fermi}-LAT, similar to
  the spectral energy distribution observed for the Crab
  pulsar.
\end{abstract}


\section{INTRODUCTION}

Millisecond pulsars (MSPs) represent a distinct population of old
rotation-powered pulsars with short spin periods, $P\le 25$ ms, and
small spin-down rates, typically of order $\dot{P}_{i}\sim
10^{-20}$. Their timing properties imply a relatively low surface
magnetic dipole field strength of $B_{\rm
  surf}\propto(P\dot{P}_{i})^{1/2}\sim10^8-10^{10}$~G and a large
characteristic spin-down age of $\tau\equiv P/2\dot{P}_{i}\ge 1$ Gyr.
These rapidly rotating neutron stars (NSs) are spun up by accretion of
matter and angular momentum from a close stellar companion in a
low-mass X-ray binary.

Of the $\sim$300 known radio MSPs, dozens have been detected in
X-rays, and are broadly grouped into three distinct categories based
on their dominant emission mechanism: (i) pulsed thermal radiation
from the NS magnetic polar caps
\citep[e.g.,][]{Bog06,Zavlin06,For14}, (ii) unpulsed shock emission
due to a pulsar wind-driven intrabinary shock
\citep[e.g.,][]{Bog05,Bog14}, and (iii) pulsed non-thermal radiation
from the pulsar magnetosphere \citep[see, e.g.,][and references therein]{Zavlin07}.

\begin{deluxetable*}{lccccccll}
\tablecolumns{11} 
\tablewidth{0pt}  
\tablecaption{Summary of the Properties of MSPs$^1$ Observed with \nustar \label{tab:msp_summary}}  
\tablehead{
\colhead{Pulsar}  & \colhead{R.A.}  & \colhead{Decl.}  & \colhead{$d$} & \colhead{$P$} & \colhead{$\dot P$}  & \colhead{$\dot E$}        & \colhead{$B$}  & \colhead{Type}  \\
\colhead{(PSR name)}        &\colhead{(J2000)}& \colhead{(J2000)}& \colhead{(kpc)}& \colhead{(ms)} & \colhead{}          & \colhead{(erg s$^{-1}$)}  & \colhead{(G)}     & \colhead{}}
\startdata                       
%
\bf J0218$+$4232 & 02 18 06.3 & $+$42 32 17.3 &  3.15 & 2.323 & $7.74\times 10^{-20}$ & $2.44\times 10^{35}$ & $4.29\times 10^{8}$  & {\bf Magnetospheric}   \\   
J0437$-$4715     & 04 37 15.8 & $-$47 15 09.1 &  0.16 & 5.757 & $5.73\times 10^{-20}$ & $1.19\times 10^{34}$ & $5.81\times 10^{8}$  & Thermal  \\                 
J1023$+$0038     & 10 23 47.6 & $+$00 38 40.8 &  1.37 & 1.688 & $1.20\times 10^{-20}$ & $9.85\times 10^{34}$ & $1.44\times 10^{8}$  & intrabinary shock    \\    
J1227$-$4853     & 12 27 58.7 & $-$48 53 42.7 &  1.80 & 1.686 & $1.11\times 10^{-20}$ & $9.13\times 10^{34}$ & $1.38\times 10^{8}$  & intrabinary shock  \\      
J1723$-$2837     & 17 23 23.1 & $-$28 37 57.1 &  0.72 & 1.856 & $7.54\times 10^{-21}$ & $4.66\times 10^{34}$ & $1.20\times 10^{8}$  & intrabinary shock  \\      
\bf B1821$-$24 & 18 24 32.0 & $-$24 52 10.8 &  5.50 & 3.054 & $1.62\times 10^{-18}$ & $2.24\times 10^{36}$ & $2.25\times 10^{9}$  & {\bf Magnetospheric}  \\    
\bf B1937$+$21   & 19 39 38.5 & $+$21 34 59.1 &  3.50 & 1.558 & $1.05\times 10^{-19}$ & $1.10\times 10^{36}$ & $4.09\times 10^{8}$  & {\bf Magnetospheric} \\     
J2129$-$0429     & 21 29 45.0 & $-$04 29 05.5 &  1.10 & 7.620 & $2.60\times 10^{-19}$ & $3.90\times 10^{34}$ & $1.42\times 10^{9}$  & intrabinary shock \\      
J2339$-$0533     & 23 39 38.7 & $-$05 33 05.3 &  1.10 & 2.884 & $1.41\times 10^{-20}$ & $2.32\times 10^{34}$ & $2.04\times 10^{8}$  & intrabinary shock         
\enddata

\tablenotetext{}{{\bf Note}. Compiled from the ATNF radio pulsar catalog
  \citep{man05}. 
Provenance of coordinates used for the barycenter corrections is given in the text. 
The \nustar\ observations of the center of 47
  Tuc, which contains at least 25 MSPs \citep{Pan16}, are not included
  in this Table; X-ray emission from the MSPs in 47 Tuc is completely
  dominated by the bright overlapping LMXB X9 \citep{Bahramian17} and
  are not detectable.}
\label{tab:msp_summary}
\end{deluxetable*}

The MSPs in the latter category provide an important testbed for
theoretical models of pulsar electrodynamics and emission physics
because they exhibit remarkably similar properties to non-recycled
young pulsars like the Crab, this despite their faster spin (and as a
result, much lower altitude of the pulsar light cylinder) and weaker
surface magnetic fields ($\sim$$10^8-10^9$~G). This is likely a
consequence of their high magnetic fields in the vicinity of the light
cylinder ($\sim$$10^5$ G), where the non-thermal radiation is thought
to originate. These MSPs are also Crab-like in the sense that they are
known to exhibit giant radio pulses
\citep{Cognard96,Romani01,Knight06}, which in the case of PSR
J0218$+$4232 appear to be closely aligned in spin phase with the two
soft X-ray pulses.  Constraining the X-ray emission properties over as
wide a an photon energy range as possible may offer vital clues
regarding the emission physics of MSPs, and by extension, all pulsars.
Specifically, important insight into the detailed physics of MSP
magnetospheric emission can be gained by investigating the energy
dependence of the pulsations and the broadband high-energy (soft
X-rays through $\gamma$-rays) spectral properties of these pulsars.


In this paper, we present the first \textit{Nuclear Spectroscopic Telescope Array}
  (\textit{NuSTAR}) hard X-ray (3$-$79~keV)
observations of three magnetosphericially driven MSPs, PSRs B1821$-$24,
B1937$+$21, and J0218$+$4232, whose properties are given in
Table~\ref{tab:msp_summary}.  These are among the most energetic MSPs
known and the only ones in their category sufficiently bright to
measure above $10$~keV. The \nustar\ data allow us to study their
X-ray spectra in an energy range beyond what was previously possible
and to consider their broader spectral energy distribution (SED) with
respect to \fermi\ Large Area Telescope (LAT) spectra.  To fully exploit the phase information
for these rapidly spinning pulsars (1.6$-$3.0~ms) we devise a method
to self-correct the data for drift and fluctuations of the \nustar\
clock, whose short-term timing variance is otherwise on the order of
the pulse periods for all three pulsars.

This work is laid out as follows. In \S 2, we present our \nustar\
observation of \psr, along with a spectroscopic reanalysis of archival
\xte\ and \chandra\ data. These observations and their reductions are
detailed in \S2.1.  In \S2.2 we explain our technique used to correct
the photon arrival time for \nustar\ clock inaccuracies. The resulting
timing, image, and spectral analyses for \psr\ is presented in \S2.3-2.5, 
respectively. In \S3, we present spectral analysis of
archival \nustar, \chandra, and \xmm\ data for PSR B1937$+$21; the
results are given in \S3.1.  Similarly, in \S4 we present the
observations and results of a spectroscopic analysis of PSR
J0218$+$4232 using existing \nustar\ and \xmm\ data.  In \S5, we
discuss the SED of the three MSPs and offer conclusions in \S6.

\section{PSR B1821$-$24}

The bright 3.05~ms pulsar \psr\ (also known as PSR J1824$-$2452A)
resides in the globular cluster M28 and has the distinction of being
the most energetic MSP known, with a spin-down luminosity\footnote{The
  measured spin-down rate $\dot{P}$ and derived spin-down luminosity
  $\dot{E}\propto \dot{P}/P^3$ are not significantly affected by
  the acceleration of the pulsar in the globular cluster potential
  \citep[see][]{Johnson13}.}  of $\dot{E}=2.2\times10^{36}$~ergs~s$^{-1}$. It was the first MSP discovered in a globular cluster
\citep{Lyne87} and the first from which non-thermal pulsed X-rays were
detected \citep[using \textit{ASCA};][]{Saito97}.  Nearly all of the
X-ray emission from \psr\ is contained within two exceptionally narrow
pulse peaks, with a duty cycle of only a few percent \citep{Rut04,Ray08}.

The non-thermal spectrum of \psr\ is well-characterized by an absorbed
power-law model with photon index $\Gamma\approx1.3$ yielding a
luminosity of $L($0.3-8$ \ {\rm keV}; 5.5 \ {\rm kpc} ) =
1.4\times10^{33}\Theta$ erg~s$^{-1}$, where $\Theta$ is an unknown
beaming fraction of the X-ray emission pattern \citep{Rots98,Bog11}.
In this sense, \psr\ is more similar to the young, energetic pulsars
than to a typical MSP. Measurements with \xte, sensitive in the hard
X-ray band (2$-$100~keV), are only able to detect pulsations up to
$\sim$13~keV, despite deep observations \citep[see, e.g.,
][]{Rots98,Ray08}. Whether this is an intrinsic cutoff on the pulsar
flux or an instrumental limitation is unknown.

In the following subsections, we present \nustar\ observations of \psr,
analyzed together with the accumulated \chandra\ and \xte\ data, which
allows us to extend the previous spectral results up to $79$~keV.

\subsection{OBSERVATIONS AND ANALYSIS}

\subsubsection{\nustar}

We observed \psr\ with \nustar\ on 2015 June 6 (ObsId 30101053002)
and on 2015 June 21 (ObsId 30101053004), as part of the AO1 Guest Observer
program.  \nustar\ consists of two co-aligned X-ray telescopes, with
the corresponding focal plane modules FPMA and FPMB that provide
$18^{\prime\prime}$ FWHM imaging resolution over a 3$-$79~keV X-ray
band with a characteristic spectral resolution of 400 eV FWHM at 10
keV \citep{Harrison2013}.  The reconstructed \nustar\ coordinates are
accurate to $7\farcs5$ at the 90\% confidence level. The nominal
timing accuracy of \nustar\ is $\sim$2~ms rms, after correcting for
drift of the onboard clock, with the absolute timescale shown to be
better than $< $$3$~ms \citep{Mori14, Madsen15}.  Although this is
insufficient to resolve the 3.05~ms period of PSR~B1821$-$24, we
devise a method, presented herein, to correct the photon arrival times
to better than $15\ \mu$s accuracy.

\nustar\ data were reprocessed and analyzed using {\tt FTOOLS}
09May2016\_V6.19 ({\tt NUSTARDAS} 14Apr16\_V1.6.0) with the \nustar\
Calibration Database (CALDB) files of 2016 July 6.  The resulting data
set provides a total of 140.7~ks and 55~ks of net good exposure time
for the two pointings, respectively. For the timing analysis, the
nominal \nustar\ clock correction was applied to the photon arrival
times which are converted to barycentric dynamical time (TDB) using the
DE405 solar system ephemeris and radio timing coordinates of
\cite{Ray08}. For all subsequent analysis we merged data from both
FPM detectors.


Although the globular cluster M28 hosts at least 11 other pulsars
\citep{Beg06}, they are all over a factor of $\sim$100 fainter
than \psr\ in X-rays \citep{Bog11} and thus cannot be studied
effectively with \nustar.

\subsubsection{\xte}

\psr\ has been observed extensively in hard X-rays with \xte\ during
the mission lifetime, primarily for the purpose of absolute spacecraft
time calibration and to measure the X-ray-to-radio phase lags
\citep[][]{Rots98,Ray08}. In anticipation of the \nustar\ observations,
we collectively analyzed all available archived \xte\ data on \psr\
for the first time. A total of 180~ks of exposure times were was accumulated
during three programs: P20159 (1997 Feb 12-13, 16.9~ks), P90053 (2005
Feb 28-Mar 1, 14.9 ks), P92008 (2006 Jul 6-8, 48.4~ks ; 2006 Oct
9-10,51.2~ks ; 2007 Apr 25-26, 48.5~ks).  These observations were made
with the Proportional Counter Array (PCA; Jahoda et al. 1996) onboard
\xte.  The PCA consists of five collimated xenon proportional counter
units (PCUs), each having a front propane anticoincidence layer, with
a total effective area of $\sim$$6500$~cm$^{-2}$ over a $1^{\circ}$
field of view (FWHM). Each PCU is sensitive to photons in the energy
range of $2-60$~keV, with an energy resolution of 18\% at 6 keV
recorded in 256 channels.

Data were collected in {\it GoodXenonwithPropane} mode with two to five PCU
modules in use at any given time, with an average of 3.7 PCUs active
overall. In this mode, photon arrival times are recorded with $1\mu s$
resolution. We analyzed events from the top xenon layers of each PCU
only. Including the second layers resulted in no significant
improvement in the signal-to-noise ratio. The data were time-filtered
using standard criteria and analyzed using the \textit{RXTE} FTOOLS
package.

\xte\ PCA phase-resolved light curves and spectra for \psr\ were
constructed using {\tt fasebin} and related FTOOLs. By folding on the
radio ephemeris of \cite{Ray08} we obtained a summed spectrum in 100
phase bins including data from the 22 observation segments. Energy-selected 
light curves were generated using {\tt fbssum} and a combined
spectrum comprising data from all \xte\ programs were added using {\tt
  fbadd}. Appropriate response matrices were created for each PCU using
{\tt pcarsp} for the phased-average spectra extracted using {\tt
  seextrct}.

\subsubsection{\chandra}

In our analysis, we also make use of archival \chandra\ ACIS-S imaging
and spectroscopic data of M28 from 2002 (ObsIDs 2684, 2685, 2683) and
2008 (ObsIds 9132, 9133) totaling 240 ks and previously presented in
\citet{Beck03}, \citet{Bog11}, and \citet{Serv12}. We also consider
\chandra\ HRC-S observations of \psr\ obtained in timing mode (ObsIDs
2797 and 6769), which offer 16 $\mu$s time resolution. 
The HRC is most sensitive to photons below 2 keV but provides no
useful spectral information.  The data reduction and analysis
procedures of all \chandra\ observations are detailed in
\citet{Bog11}.

\begin{figure}[t]
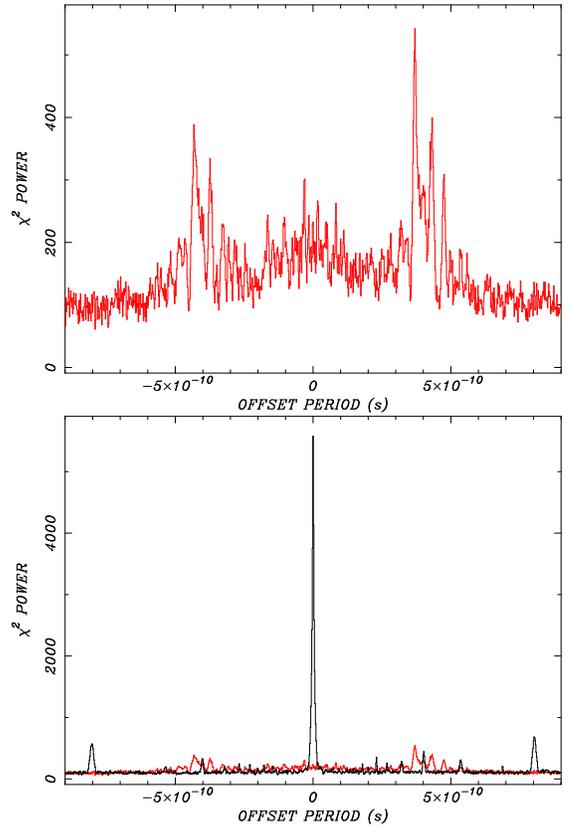

\centering
\psfig{figure=fig1a.ps,height=0.85\linewidth,angle=-90} 
\psfig{figure=fig1b.ps,height=0.85\linewidth,angle=-90} 
\caption{\nustar\ periodogram of \psr\ folded around the expected
  3.05~ms spin period. {Top:} result obtained for the nominal
  processed data set.  Power in the pulsed signal is spread out over a
  number of ``sidebands'' around the expected period.
  {Bottom:} the result obtained using the same data set corrected
  for timing inaccuracies, as described in \S\ref{sec:method}
  ({black}).  The periodogram of the uncorrected data ({red})
  from the upper panel is reproduced for reference.}
\label{fig:nustarclockerr}
\end{figure}

\subsection{Correcting the \nustar\ Photon Arrival Times\label{sec:method} }

\begin{figure}[t] 
\psfig{figure=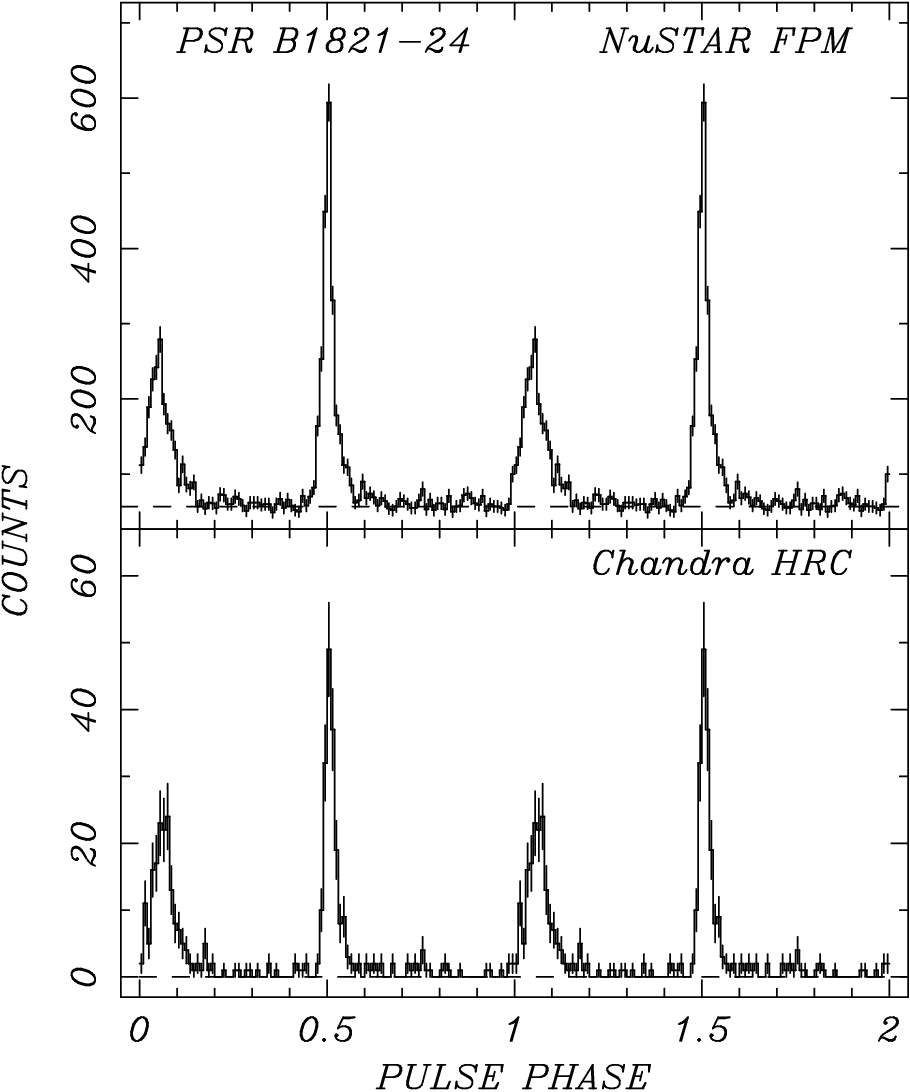,height=0.85\linewidth,angle=-90} 

\caption {Comparison of the \nustar\ ({Top}) and \chandra\
({Bottom}) pulse profiles of the 3.05~ms pulsar \psr.  The
\nustar\ 3$-$79~keV data are corrected for timing inaccuracies, as
described in \S\ref{sec:method} and the \chandra\ HRC data were obtained
in timing mode.  The two data sets complement each other, with the HRC
data dominant below 2~keV and the \nustar\ data above 3~keV. The
background level is indicated by the dashed line. The profiles
contains 100 phase bins and phase zero is arbitrarily aligned so that
the main pulse peak falls on $\phi=0.5$. Two cycles are shown for
clarity.}
\label{fig:b1821_lightcurve} 
\end{figure}
 
Before we can take advantage of the millisecond time resolution
available with the \nustar\ FPM imagers we must address several clock
issues relevant to observing MSPs.  The \nustar\ clock frequency is
known to drift at a steady rate over time and is typically adjusted
every few days by means of a coarse divisor to limit the wandering
from a nominal 24~MHz rate. A model of the clock drift is stored in
the clock-correction file and used by the FTOOL {\tt barycorr} to
correct the \nustar\ photon arrival times in conjunction with the
barycenter correction. However, unaccounted-for root mean square
residuals of order $\sim$2~ms remain in the clock drift model.

In addition, during telemetry downlinks (several times a day) the 
spacecraft times are adjusted to the coordinated universal time (UTC) system.
This can introduce a phase shift due to an uncalibrated $\sim$1~ms
relative offset in the absolute time stamps between ground stations 
(C.~Markwardt, private communication).  To circumvent the latter potential problem,
all data for the \psr\ observations were acquired using a single
(Malindi) ground station, as was used for the other two pulsars
discussed herein. A complete description of the \nustar\ event time
assignment can be found in \cite{Madsen15}.

Most consequential for fast pulsar timing is the variability of the
clock rate on orbital timescales ($97$~minutes).  Our investigation of the
\nustar\ timing data reveals that the clock is stable to better than
at least $15 \ \mu s$ during the spacecraft night, when the solar
illumination of the spacecraft is Earth-blocked.  However, during
spacecraft day, the clock evidently wanders sufficiently between
orbit-to-orbit observing intervals to lose phase cycle count for a
millisecond pulsar.  Thus, it is not possible to recover a coherently
pulsed signal over several orbits.  This is best illustrated by a
pulsar search for \psr\ around the expected frequency using the
nominal \nustar\ processing, including barycenter and clock
corrections.  As shown in the top panel of
Figure~\ref{fig:nustarclockerr}, the search periodogram yields a
complex power spectrum of low significance with a multitude of
``sidebands'' spanning $\delta f \approx 9\times 10^{-5}$~Hz.

Fortunately, the observations of \psr\ were taken exclusively during
intervals of Earth-block and its pulsed signal is sufficiently strong
to measure a pulse frequency and phase over a single orbit.  For the
60 orbits that constitute the full data set, on average 75 photons per
orbit were collected in the source aperture for a typical 3.3~ks of
exposure. Detecting a signal with such few photons is uniquely
possible because of the very sharp, asymmetric double peak profile,
with $\approx$100\% modulation and very low detector background.  We
use this fact to correct errors in the frequency and phase introduced
between orbits.

Our method is to break up the observation into orbit-sized segments
and correct for the advanced or retarded clock rate by adjusting the
photon arrival times in the following sense,

$$ t^{\prime} = t \times \alpha_i  + \beta_i $$

\noindent where, for the $i{\rm th}$ segment, $\alpha_i$ is the ratio of
the observed frequency of the pulsar relative to the expected
frequency, and $\beta_i$ is the phase offset at the adjusted times
relative to the expected phase. The scale factors $\alpha$, around
$\pm 4 \times 10^{-8}$, are measured using the $\chi^2$ statistic,
optimal for the complex pulse shape. The phase offsets $\beta$ are
determined by cross-correlating the pulse profile for each 
segment with an iterated high-statistic template.  The $\beta$'s
ensure that the scaled photon arrival times for each orbit align on
the fiducial ephemeris relative to the start of the observation, with the
main peak (P1) adjusted to phase $\phi_{P1}=0.5$. This method
accounts for all unmodeled orbit-to-orbit clock shifts.

To constructed the set of 60 $\alpha$'s and $\beta$'s for \psr, we
extracted photons from an optimal combination of source aperture
($r=0.85'$) and energy range (3$-$25 keV) that maximizes the pulsar
signal and cross-correlated the folded light curves for each segment
in 100 phase bins with the template.  We applied these coefficients to
the photon arrival times to generate the corrected event files that
can now be coherently folded on the input ephemeris to recover the
full pulsar signal.\footnote{However, we cannot be sure that the phase
  count was strictly maintained. For example, if the phase count were lost,
  it may not be possible to fold coherently a second pulsar in the
  field using the corrections for \psr.} These files are used in all
subsequent phase-resolved imaging and spectral analysis presented
below.  The resulting periodgram search using the corrected data is
shown in the bottom panel of Figure~\ref{fig:nustarclockerr}.

\subsection{\nustar\ Timing Analysis}

The folded light curve, using data merged from both \nustar\ pointings
of \psr, fully resolves the peak of the pulsed signal in $30 \ \mu$s
phase bins, limited ultimately by the accuracy of the
cross-correlations, detailed above. In
Figure~\ref{fig:b1821_lightcurve}, we compare the 3$-$79 keV \nustar\
pulse profile to the \chandra\ HRC profile, obtained from a
$3^{\prime\prime}$ diameter aperture, barycentered and folded on the
ephemeris of \cite{Ray08}. Although the HRC is most sensitive to
photons below 2 keV, there is no clear energy dependence; the observed
morphology of the two profiles are very similar, showing the same
sharp, asymmetric peak, despite the non-overlapping energy bands.  The
background level, estimated from a circular annulus accounts for
nearly all of the off-pulse flux. A small contribution from the
unresolved M28 X-ray sources in the \nustar\ point-spread function
(PSF) accounts for the rest, as discussed below. 

\begin{figure*} [t]
\centering
\psfig{figure=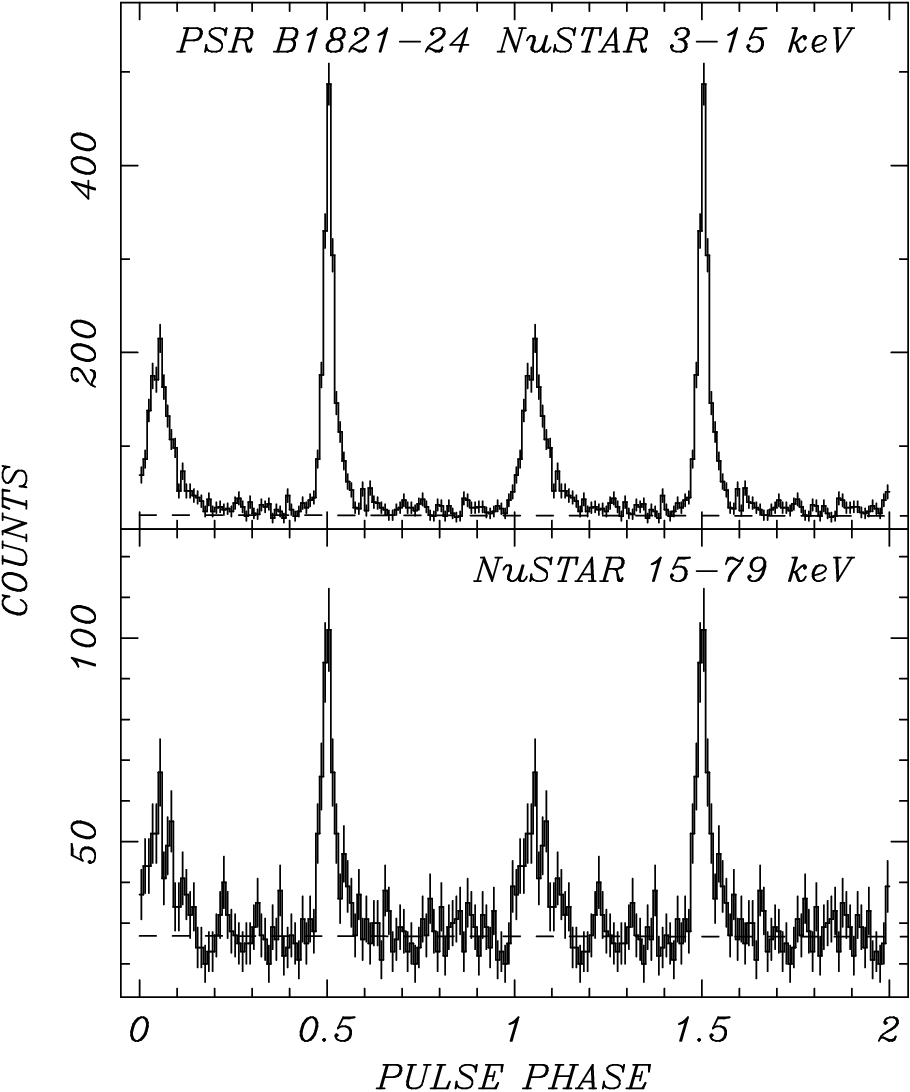,height=0.4\linewidth,angle=-90} ~~~~
\psfig{figure=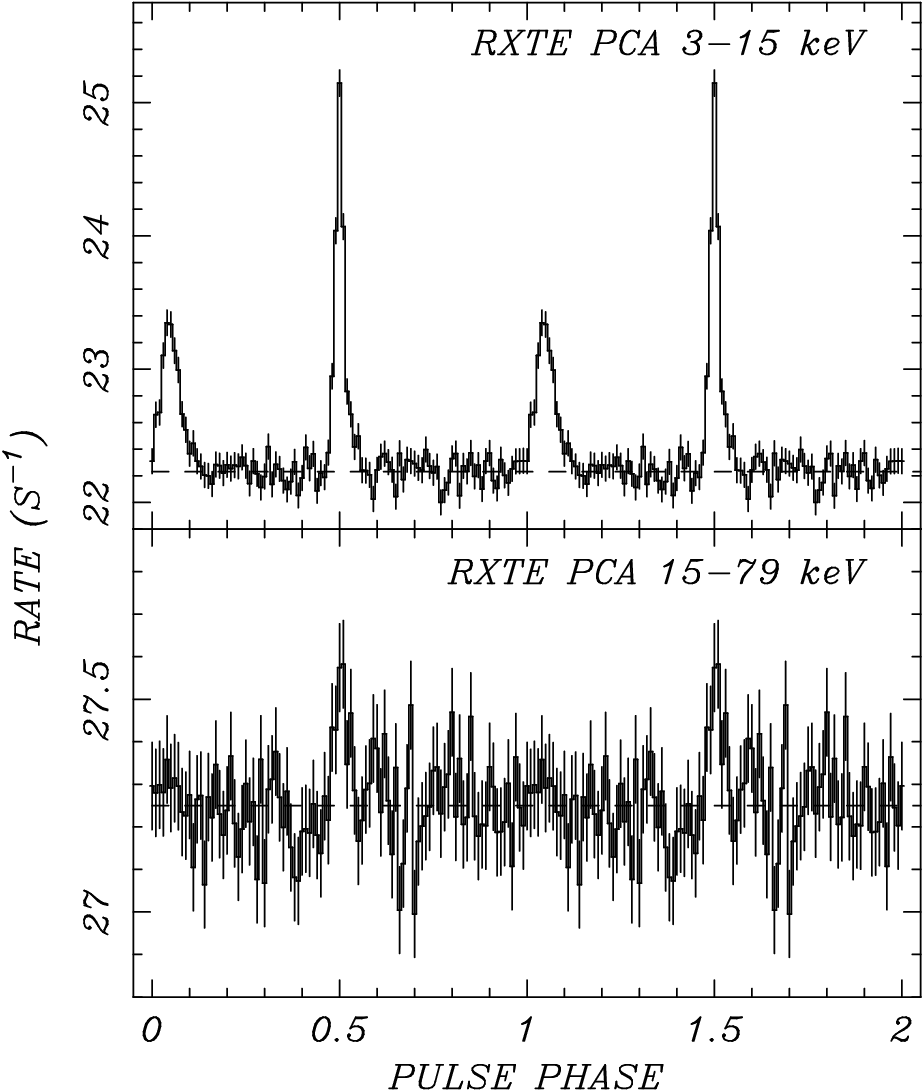,height=0.405\linewidth,angle=-90} 
\caption{Comparison of the \nustar\ ({left}) and \xte\
  ({right}) pulse profiles of \psr\ in two energy bands. The
  \nustar\ photon arrival times have been corrected for timing
  inaccuracies, as described in \S\ref{sec:method}, and the \xte\ PCA
  data are from the top layer only. The background level is indicated
  by the dashed line. These profiles contains 100 phase bins and phase
  zero is arbitrarily aligned so that the main pulse peak falls on
  $\phi=0.5$. Two rotational cycles are shown for clarity.}
\label{fig:b1821_xte_lightcurve} 
\end{figure*}

Figure~\ref{fig:b1821_xte_lightcurve} also compares the \nustar\ and
\xte\ pulse profiles in two adjacent energy bands, 3$-$15 and
15$-$79~keV. At the lower energy, the profiles from the two missions
are again strikingly similar.  However, above 15~keV the signal in the
\xte\ data is barely visible above the background, while the \nustar\
pulsations are clearly detected with high signal-to-noise ratio, up to
$\sim$30 keV. This pulse signal is evidently not cut off below 15~keV,
but it is instead lost in the high background of the non-imaging \xte.  These
light curves illustrate that the timing accuracy of the corrected
\nustar\ data set can be on par with the highest quality \xte\ and
\chandra\ data.

\begin{figure*}[!t]
\centering
\psfig{figure=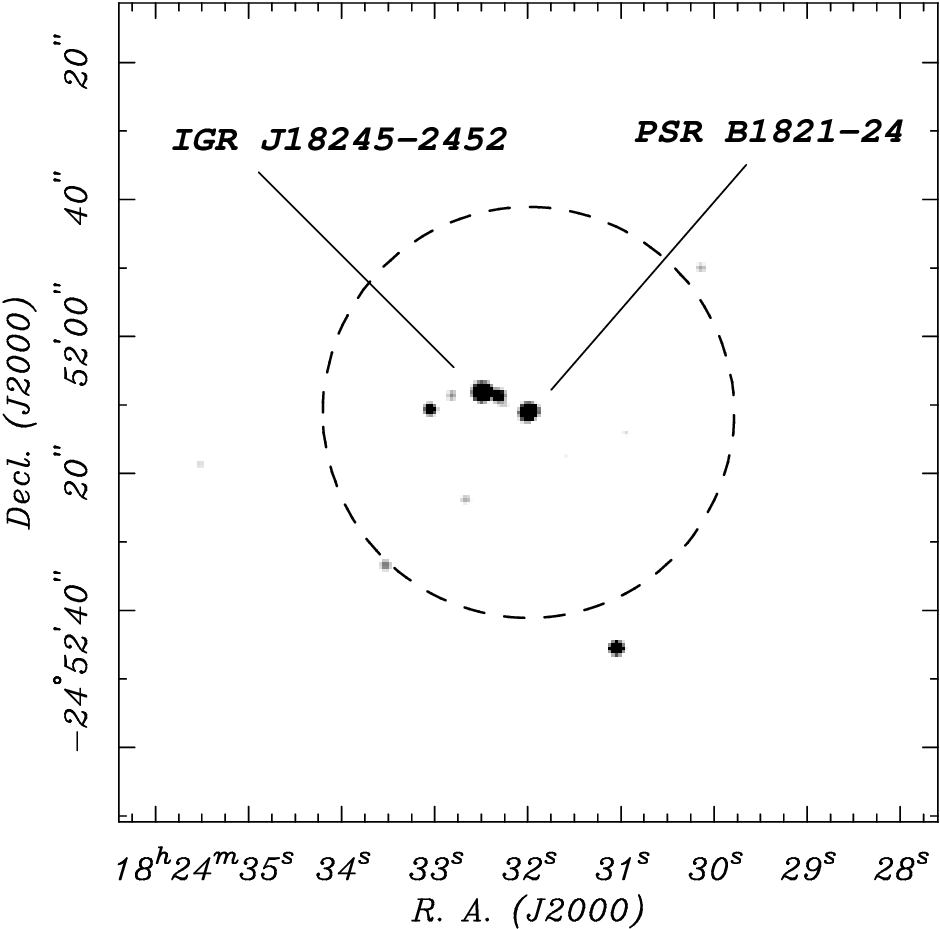,height=0.45\linewidth,angle=-90}~~~~~
\psfig{figure=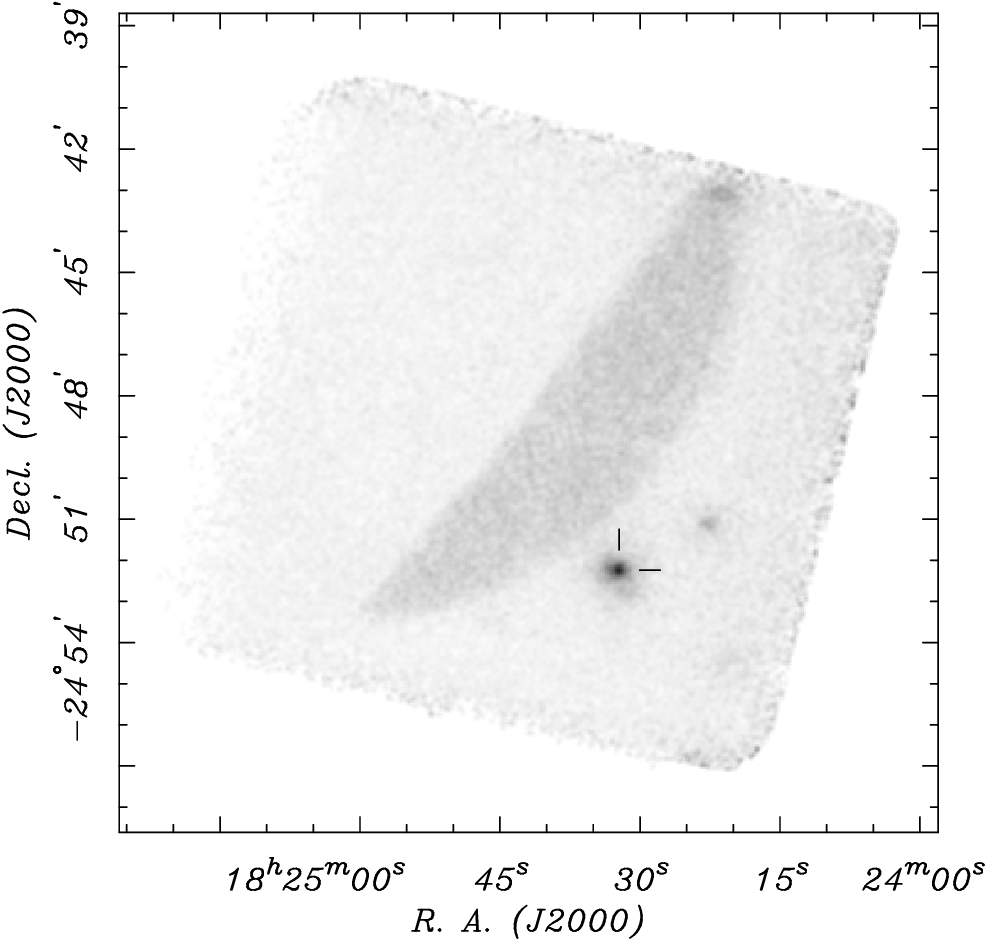,height=0.45\linewidth,angle=-90}\\
\vfill
\psfig{figure=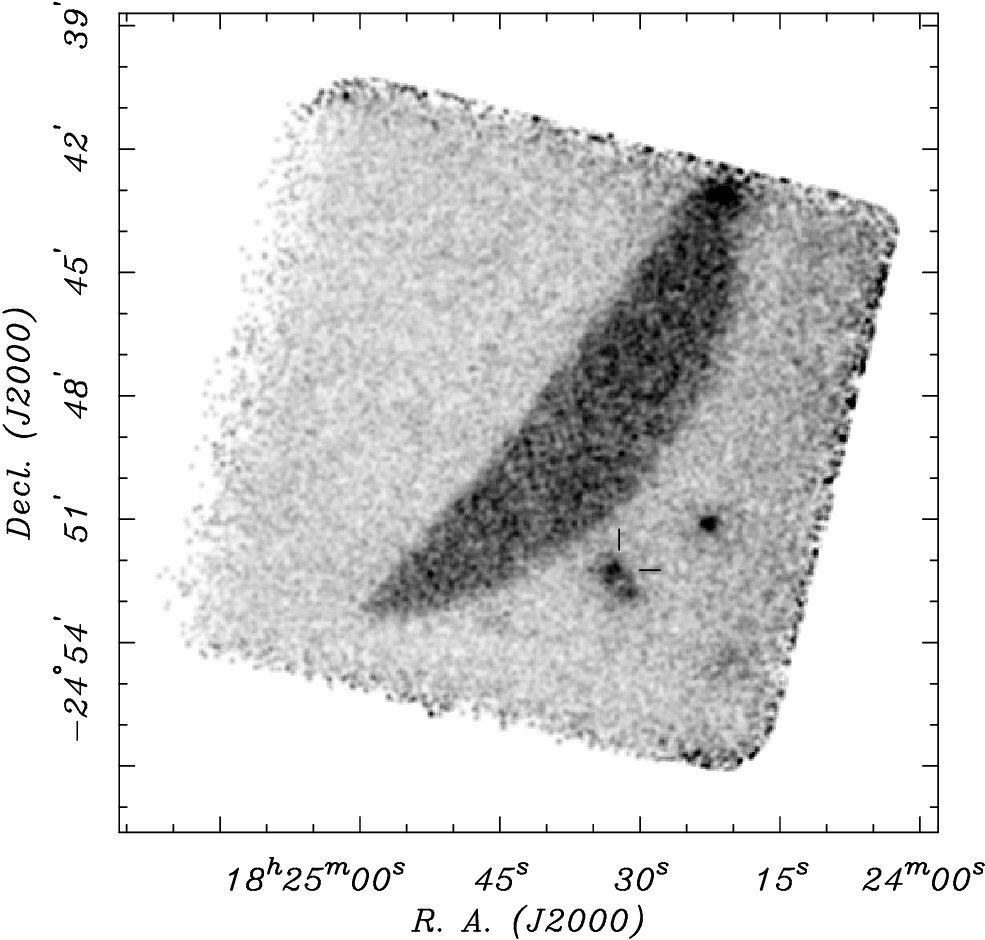,height=0.45\linewidth,angle=-90}~~~~~
\psfig{figure=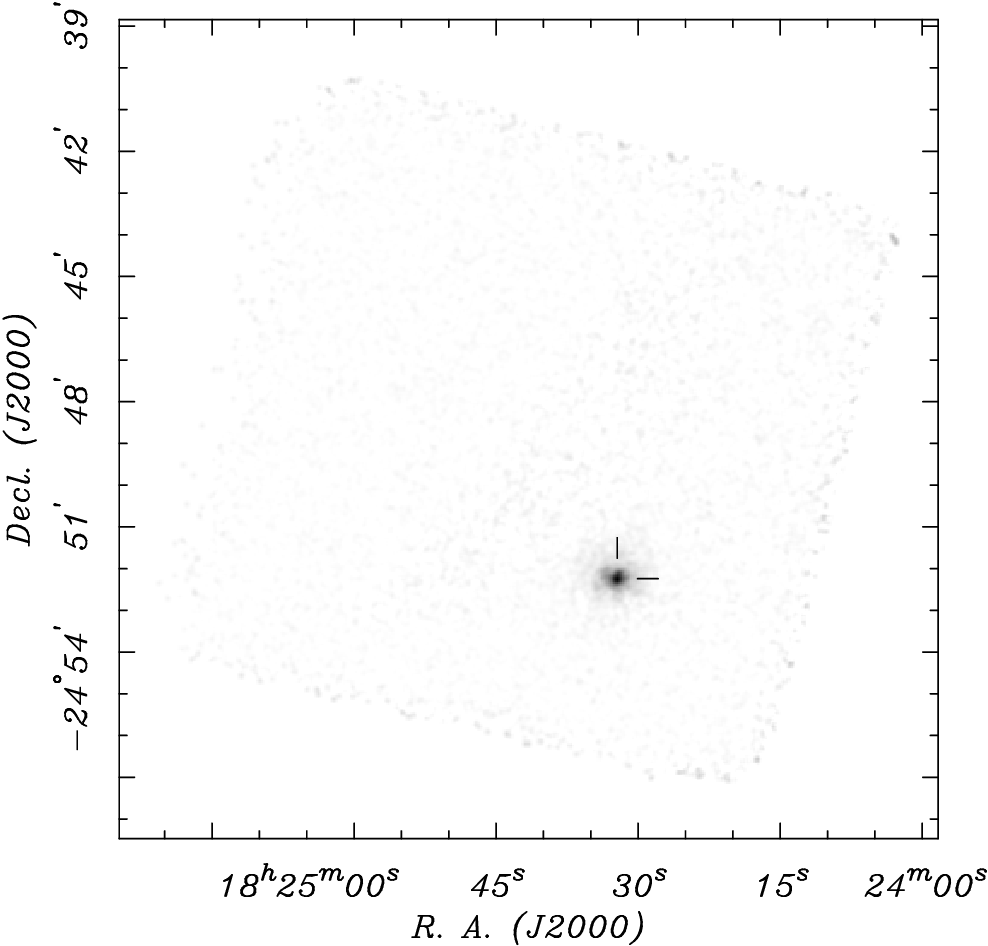,height=0.45\linewidth,angle=-90}
\caption{{Top left ---} A reference high-resolution \chandra\
  ACIS-S 3$-$10 kev image of the core of M28. The dashed circle of
  radius $0.5\farcm$ centered on \psr\ corresponds to the $\sim$50\%
  enclosed fraction of the \nustar\ point-spread function. In this
  observation from 2008, the transient source IGR~J18245$-$2452
  \citep{Pap13} was more luminous than \psr\ but was $\sim$2 orders of
  magnitude fainter at the time of the \nustar\ observations in 2015.
  Above $\sim$3 keV, \psr\ is by far the brightest persistent source
  in M28. {Other three panels ---} \nustar\ exposure-corrected and
  smoothed 3$-$20~keV X-ray images of the field containing \psr.  {\it
    Top right ---} The pulse phase-averaged image showing the pulsar
  and source \#4 from \citet{Beck03}. Also evident is stray light
  contamination from the luminous X-ray binary GS 1826$-$24.  {
    Bottom left ---} The off-pulsar image that shows evidence of
  contamination from unresolved hard sources in M28 coincident with
  \psr. {Bottom right ---} The pulsar image generated by
  subtracting the scaled off-pulse image from the phase-averaged image.}
\label{fig:images}
\end{figure*}

\begin{figure}[] 
\centerline{ 
\hfill
\psfig{figure=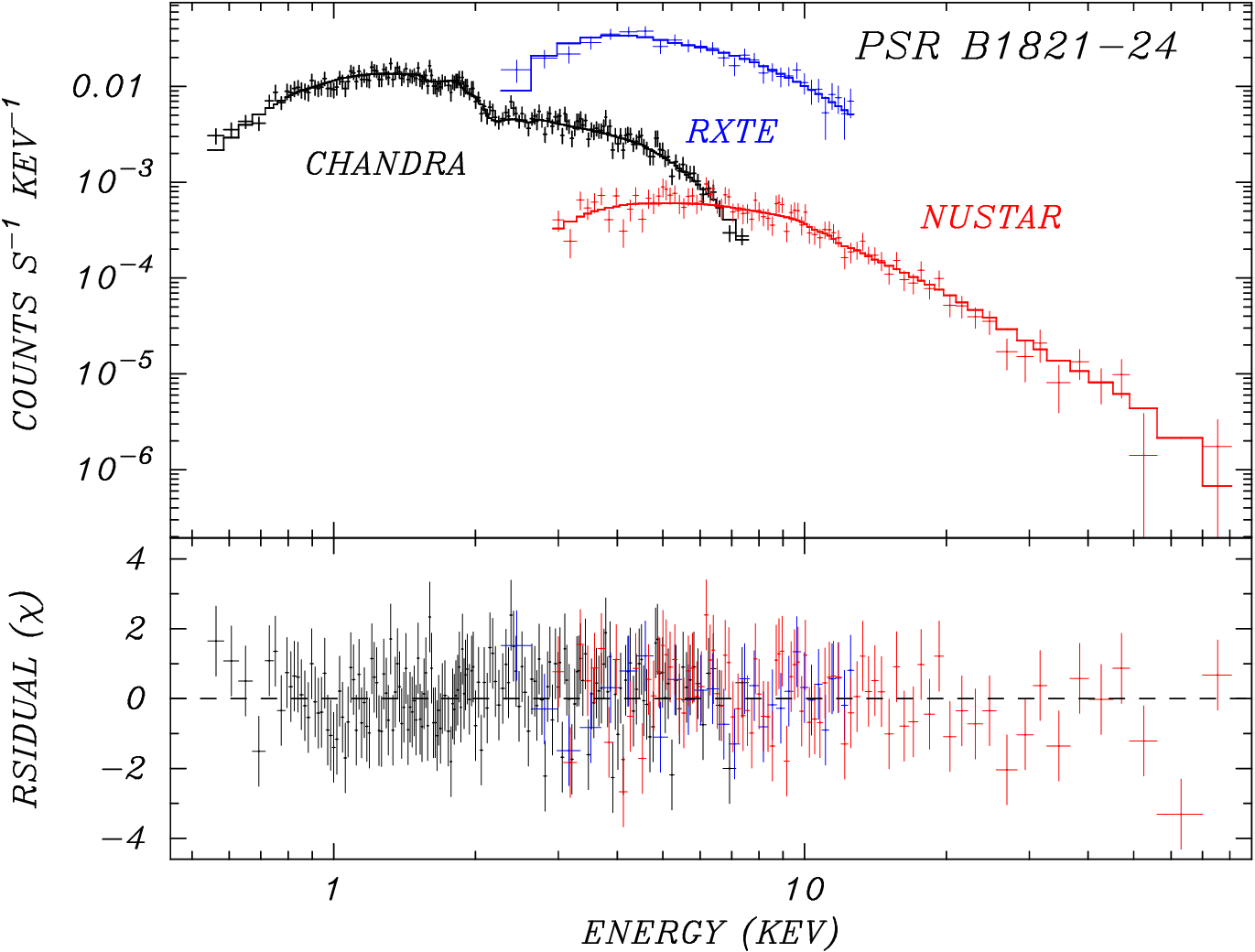,height=0.85\linewidth,angle=-90}
\hfill 
} 
\caption{Phase-resolved  \nustar\ and \xte\ spectra of \psr\ fitted
  simultaneously with the phase-averaged \chandra\  spectrum to an
  absorbed power-law model with independent normalizations.  The upper
  panel presents the \chandra\ ({black}), \xte\ ({blue}), and
  \nustar\ ({red}) spectral data points ({crosses}) along with
  the best-fit model ({solid lines}) given in
  Table~\ref{tab:spectable}.  The lower panel shows the fit residuals
  in units of sigma.}
\label{fig:b1821_chandra_nustar_spec}
\end{figure}

\subsection{\nustar\ Image Analysis}

\psr\ is situated in the dense core of M28 where numerous other X-ray
sources are present, as seen in the sub-arcsecond angular resolution
\chandra\ images \citep{Beck03,Bog11}. To the \nustar\ imaging resolution, the M28
cluster core is unresolved.  This does not pose a problem since \psr\
is the brightest source above $\sim$5 keV.  Indeed, the bright hard
X-ray source in M28 has a position fully consistent with
that of \psr.  The transient IGR~J1824$-$24525 in M28 occasionally
reaches comparable or higher luminosities and exhibits a hard spectrum
\citep[][]{Pap13,Lin14a}. However, this object spends the vast
majority of the time in a dormant state, with luminosity
$\sim$$10^{31}$~erg~s$^{-1}$ so its quiescent emission in the hard
X-ray band during the \nustar\ exposure is negligible for all
practical purposes. 

Figure~\ref{fig:images} presents 3$-$20~keV exposure-corrected images
of the \nustar\ M28 field smoothed using a
$\sigma=3.\!^{\prime\prime}7$ Gaussian kernel and scaled linearly. To
look for possible unpulsed X-ray emission from \psr\ we examined the
\nustar\ phase-average, on-pulse, and off-pulse images. We define the
phase range corresponding to the on-pulse interval to include photons from
both pulse peaks, P1 (10 of 100 phase bins) and P2 (16 of 100 phase
bins), with the off-pulse interval containing the remaining phase bins
(see Figure~\ref{fig:b1821_lightcurve}).

The phase-averaged \nustar\ image shows the bright pulsar as well as
the emission coincident with \chandra\ source \#4 reported by
\citet{Beck03}, likely an active galactic nucleus or a cataclysmic
variable, to the northwest of \psr.  Also evident in the image is a
large bright crescent of emission that corresponds to contamination
from single-bounce photons off the \nustar\ optics originating from
the bright X-ray binary GS 1826$-$24 \citep{Ubertini99}, which is
situated 1.6$^{\circ}$ from the core of M28. From the off-pulse image
it is clear that there is detectable emission at the location of the
pulsar after accounting for the off-source background. This remainder
can be attributed to hard X-ray emission from the numerous other
sources in M28 that are unresolved by \nustar. We estimate this
component from the scaled \chandra\ flux in the 3--10 keV band that
fall within the \nustar\ PSF. Its contribution is fully consistent
with the off-pulse component of the light curve in the same band,
after allowing for the local \nustar\ background. Also evident in the
off-pulse image is an object just to the southwest of the pulsar
consistent with the position of \chandra\ source \#17 identified by
\citet{Beck03}, which is a candidate cataclysmic variable. The final
panel of Figure~\ref{fig:images} presents the off-pulse-subtracted
phase-averaged image, providing a clean representation of the \nustar\
PSF of the pulsar, as expected.

\subsection{X-Ray Spectroscopy}

In all of the following spectral analyses, extracted spectra, grouped into
appropriate channels, were fitted to an absorbed power-law spectrum
model using the {\tt XSPEC} (v12.8.2) package \citep{Arnaud96}.  All
spectral fits use the {\tt TBabs} absorption model in {\tt XSPEC} with
the {\tt wilm} solar abundances \citep{Wilms2000} and the {\tt vern}
photoionization cross-section \citep{Verner96}.

\subsubsection{\xte\ and \chandra\ Spectroscopy}

To isolate the purely pulsed emission for the \xte\ data, we used the
scaled off-pulse phase intervals for a background spectrum. This
provides a perfect representation of the non-pulsed emission. The
resulting background-subtracted \xte\ spectrum of \psr\ is then fit
simultaneously with the 240~ks \chandra\ ACIS-S3 (spatially isolated
but phase-averaged) pulsar spectrum presented in \citet{Bog11}, with
their normalizations unlinked.

The available \xte\ data contain no useful spectral information above
$\sim$13 keV. This is not necessarily due to the background, which
exceeds the source rate by an order of magnitude at all energies, but
rather for the lack of source photons and the large uncertainty in
subtracting two comparably large numbers.  With the column density
fixed to the \chandra-derived value of $N_H = 2.4\times
10^{21}$~cm$^{-2}$ \citep{Bog11}, the best-fit ($\chi^2 = 0.88$ for
198 d.o.f.),  power-law model in the $<15$~keV band yields a photon
index of $\Gamma = 1.25^{+0.12}_{-0.14}$, in agreement with \chandra\
results at lower energies. There is no evidence for flux variability.
This result represents the best X-ray spectroscopy available for \psr\
prior to the \nustar\ result presented below.

\subsubsection{\nustar\ Spectroscopy}

We can improve on the \xte\ result by taking advantage of the greatly
decreased \nustar\ background to extend the spectral measurements up to
79~keV. As with the \xte\ phase-resolved spectroscopy analysis, we
isolate the \nustar\ spectrum of the pulsar into two pulse peaks and
define a clean representation of the background using the scaled
off-pulsed spectrum.  Spectra were produced using the {\tt nuproducts}
script in the \textit{NuSTAR} FTOOL package acting on the
phase-resolved data files.  The spectra are fitted using instrument
and mirror response files generated for the phase-averaged spectra. We
add the spectra derived from the two detectors and two observations to
produce our final phase-averaged, on-pulse, and off-pulse spectral and
weighted response files.

The \chandra\ and \textit{NuSTAR} spectra were both binned with a
minimum of 30 counts per channel and fitted simultaneously with the
\rxte\ data to an absorbed power-law model with independent
normalizations.  The three data sets were restricted to the 0.5$-$8 keV, 
3$-$50~keV, and 2$-$13 keV energy bands, respectively.  The resulting fit
yields a column density of $(4.0\pm0.4)\times 10^{21}$~cm$^{-2}$ and
a photon index of $\Gamma = 1.28 \pm 0.05$ with $\chi_{\nu}^2 = 0.91$
for 226 degrees of freedom and an absorbed \nustar\ flux in the
2$-$10~keV band of $(3.7\pm0.2)\times 10^{-13}$~erg~cm$^{-2}$~s$^{-1}$
and $(3.8\pm0.1)\times 10^{-13}$~erg~cm$^{-2}$~s$^{-1}$ for the
\chandra\ spectrum (see Figure~\ref{fig:b1821_chandra_nustar_spec}).
All quoted uncertainties are at a 90\% confidence level (C.L.). For a
distance to M28 of 5.5~kpc \citep{Serv12}, the derived fluxes imply an
X-ray luminosity of $4\times10^{33}\Theta$~erg~s$^{-1}$ in the
0.3$-$79 keV range.  There is no requirement for an additional thermal
component such as may arise due to heated polar caps.  The relative
\chandra\ and \nustar\ fluxes are in agreement at the 90\% confidence
level (C.L.). A summary of the spectral results is given in
Table~\ref{tab:spectable}.

As noted in \citet{Bog11}, there may be pileup distortions of the \chandra\
spectrum at a level of a few percent, which can artificially harden the intrinsic 
source spectrum.  Applying a pileup component to the model for the \chandra\ 
data results in $\Gamma=1.29\pm0.07$ for the joint fit, in excellent agreement 
with the joint fit without pileup. This is an indication that event  pile-up in the \chandra\ 
data do not significantly bias the results. Fitting the \nustar\ data alone with the column 
density fixed to the above value yields a marginally steeper photon index of 
$\Gamma = 1.36 \pm 0.07$, suggesting the possibility of  curvature in the spectrum, as
seen in the pulsed Crab spectrum.

We also examined the \nustar\ spectra from the individual pulses, P1
and P2, and find no significant difference in their photon indices,
$\Gamma_{P1}=1.36\pm0.09$ versus  $\Gamma_{P2}=1.33\pm0.11$, in agreement
to within 2\%, well within their uncertainties. The unabsorbed flux
ratio for the two pulses is found to be $F_{P1}/F_{P2} = 1.26$.

\section{PSR B1937$+$21}

PSR B1937$+$21 (also referred to as PSR J1939$+$2134) was the first
MSP discovered \citep{Backer82} and for 25 years the fastest
spinning pulsar known ($P=1.55$ ms). With a spin-down luminosity of
$\dot{E}=1.1\times10^{36}$ ergs~s$^{-1}$, it is the second most
energetic MSP detected. This isolated pulsar lies in the Galactic
plane at a distance of $\sim$3.5 kpc.  At X-ray energies, PSR
B1937$+$21 has been detected as a pulsed source with \textit{ASCA}
\citep{Taka01}, \textit{BeppoSAX} \citep{Nicastro04}, \textit{RXTE}
\citep{Guillemot12}, \chandra\ and \textit{XMM-Newton} \citep{Ng14}.
In hard X-rays, the pulsar has so far only been detected up to
$\approx$13 keV with \xte\ \citep{Guillemot12}. Like \psr, this MSP
shows sharp pulsations with a small duty cycle and nearly 100\% pulsed
fraction.

\subsection{Observations and Results}

We examined the 41~ks \nustar\ observation of PSR~B1937$+$21 obtained
on 2015 Aug 29 as part of the Cycle 1 Guest Observer program (ObsID
30101031002; PI Ng).  These data were extracted, reduced, and analyzed
in the same fashion as described above for \psr.  For the timing
correction, an average of 66 counts per orbit were collected during 14
segments.  The pulse signal from PSR~B1937$+$21 detected in a segment
is sufficiently strong to allow us to reproduce the method presented
in \S\ref{sec:method} to correct the photon arrival times. We used the
most current published pulsar ephemeris \citep{Desvignes16} to
barycenter and phase align the pulse profile over the course of the
\nustar\ observation.

PSR B1937$+$21 was also observed with \xmm\ on 2010 March 29 for a
total exposure of 66.9 ks (ObsID 0605370101). These data were
originally presented in \citet{Ng14}. The EPIC pn data for this
observation were collected in timing mode, which offers 30 $\mu$s time
resolution but imaging along one dimension only. This data is not used
here for spectroscopy due to known spectral issues for
this mode. 

\begin{figure} [t]
\hfill
\psfig{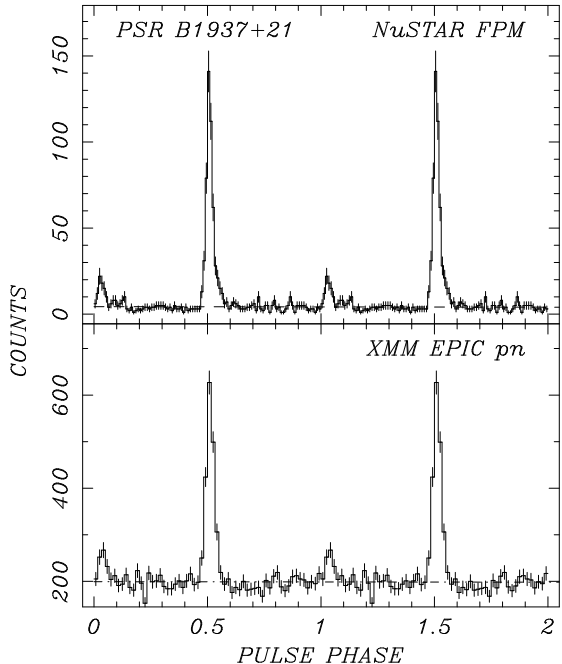} 
\hfill
\caption{Comparison of the \nustar\ ({top}) and \xmm\
  ({bottom}) pulse profiles of the 1.55~ms pulsar
  PSR~B1937$+$21.  The \nustar\ 3$-$25 keV data are corrected for
  timing inaccuracies, as described in \S\ref{sec:method}, and \xmm\
  EPIC-pn data were obtained in fast timing mode.  The background level
  is indicated by dashed line. These profiles contain 100 and 60
  phase bins for the \nustar\ and \xmm\ data, respectively.  Phase
  zero is arbitrarily aligned so that the main pulse peak falls on
  $\phi=0.5$. Two cycles are shown for clarity.}
\label{fig:b1937_lightcurve} 
\end{figure}

The folded light curves from the two missions are shown in
Figure~\ref{fig:b1937_lightcurve}. It is clear that the \nustar\ pulse
profile is fully resolved and is comparable to the best \xmm\ timing
mode observations, also shown in \citet{Ng14}. The hard X-ray pulse
profile for PSR~B1937$+$21 is qualitatively similar to that of
PSR~B1821$-$24, but with a much suppressed secondary peak. The profile
of the main peak displays the same characteristic asymmetry clearly
evident in PSR~B1821$-$24 and is detectable up to $\sim$25~keV with
\nustar, while the faint secondary pulse is lost to the background
above $\sim$8~keV. Given that the \nustar\ off-pulse emission is fully
consistent with the estimated background, we conclude that the timing
signal from PSR~B1937$+$21 is 100\% pulsed.

\begin{figure}[t]
\centerline{ 
\hfill
\psfig{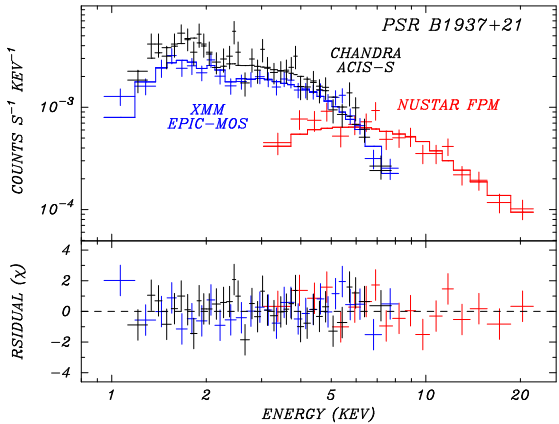}
\hfill 
} 
\caption{Phase-resolved \nustar\ spectrum of PSR B1937$+$21 fitted
  simultaneously with the phase-averaged \chandra\ and \xmm\ MOS
  spectrum to an absorbed power-law model with independent
  normalizations The upper panel presents the \chandra\ ({\it black}),
  \xmm\ ({blue}), and \nustar\ ({red}) spectral data points
  ({crosses}) along with the best-fit model ({solid lines})
  given in Table~\ref{tab:spectable}.  The lower panel shows the fit
  residuals in units of sigma.}
\label{fig:b1937_spec}
\end{figure}

For the spectroscopic analysis of PSR~B1937$+$21, we combined \nustar,
\xmm\ and \chandra\ data. Following the recipe described previously,
we extracted a \nustar\ on-pulse spectrum of PSR~B1937$+$21 and the
off-pulse spectrum to use as background.  We
extracted the \xmm\ spectrum from the EPIC MOS1/2 data following the standard procedures
as outlined in \citet{Ng14}.  We also retrieved an archival 49.5~ks
\chandra\ ACIS-S observation obtained on 2005 June 26 (ObsID 5516), which we
processed and reduced as detailed in \citet{Ng14}.
We used phase-averaged \xmm\ and \chandra\ spectra as the time
resolution of these data sets does not allow a phase-resolved spectral
analysis.

The resulting simultaneous fit to the three data sets over the
0.5$-$25~keV range is shown in Figure~\ref{fig:b1937_spec}.  Above
25~keV, the background flux dominates the \nustar\ spectrum.  The 
best-fit absorbed power-law model yields an absorbing column $N_{\rm H} =
(1.78 \pm 0.27)\times 10^{22}$~cm$^{-2}$ and spectral index $\Gamma =
1.16\pm 0.11$ with $\chi_{\nu}^2 = 0.91$ for 83 dof.  The measured
\nustar\ 2$-$10~keV unabsorbed flux is $(2.37\pm0.29) \times 10^{-13}$
erg~s$^{-1}$~cm$^{-2}$. The \xmm\ flux is 10\% higher, but consistent
with imperfect background subtraction and local contamination
\citep[see][]{Ng14}.  The \chandra\ unabsorbed flux, $(2.95\pm0.26)
\times 10^{-13}$ erg~s$^{-1}$~cm$^{-2}$, is also higher, by 20\% at
1~keV, compared to the \nustar\ value, a difference significant at the
$1.7\sigma$ level. This may be indicative of curvature in the spectrum.

\section{PSR J0218$+$4232}

PSR J0218$+$4232 is a radio luminous ~2.32~ms pulsar \citep{Navarro95}
in a two-day orbit with a helium white dwarf companion, at a distance of
$\approx$3 kpc. Its spin-down properties imply
$\dot{E}=2.4\times10^{35}$~ergs~s$^{-1}$, which places it among the
four most energetic MSPs known.  PSR J0218$+$4232 was the only MSP
marginally detected by EGRET, making it an appealing target for X-ray
observatories.  Detections in the 0.1$-$2.4 keV band with the
\textit{ROSAT} HRI and PSPC instruments were reported by
\citet{Verbunt96} and \citet{Kuiper98}. A subsequent study by
\citet{Mineo00} resulted in measured pulsed emission in the 1$-$10 keV
band in \textit{BeppoSAX} observations. Follow-up investigations have
also been conducted with \chandra\ \citep{Kuiper02} and \xmm\
\citep{Webb04}, which confirm the previous findings of two moderately
sharp pulses per period with a hard non-thermal spectrum.

PSR J0218$+$4232 is only just bright enough to correct the photon
arrival times for the clock irregularities using the method of
\S\ref{sec:method}. The pulse is not as sharp and the signal not as
strong per orbit as found for our previous MSP examples. Nevertheless,
we were able to successfully recover the pulse profile, but with
reduced timing resolution.

\subsection{Observations and Results}

\nustar\ observed PSR J0218$+$4232 during Cycle 1 on 2015 October 28
(ObsID 30101030002; PI Ng) resulting in a 34.7~ks effective exposure.
The data for this pulsar were extracted and processed in the same way
as the other two pulsars. We used the binary ephemeris from
\citet{Abdo10} to further correct the barycentered photon arrival
times for the binary orbit modulation before correcting for the clock drift.

We also retrieved archival \chandra\ data of PSR J0218$+$4232,
consisting of a 72.1 ks HRC-S exposure from 2001 November 27 (ObsID
1853) first analyzed by \citet{Kuiper02}.  Source photons were
extracted from a $3^{\prime\prime}$ diameter aperture and barycentered
using the same radio ephemeris as used for the \nustar\ data.

\begin{figure}
\centerline{ 
\hfill
\psfig{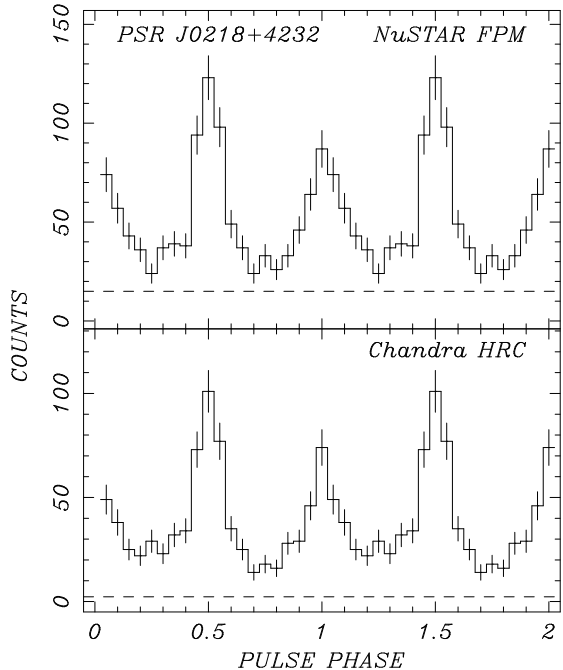}
\hfill
}
\caption{Comparison of the \nustar\ ({top}) and \chandra\
  ({bottom}) pulse profiles of the 2.32~ms pulsar
  PSR~J0218$+$4232.  The \nustar\ 3$-$25 keV data are corrected for
  timing inaccuracies, as described in \S\ref{sec:method}, and
  \chandra\ HRC-S data were obtained in fast timing mode. The
  background level is indicated by the dashed line. These profiles
  contain 20 phase bins and phase zero is arbitrarily aligned so that
  the main pulse peak falls on $\phi=0.5$. Two cycles are shown for
  clarity.}
\label{fig:j0218_lightcurve}
\end{figure}

\begin{figure} 
\centerline{
\hfill
    \psfig{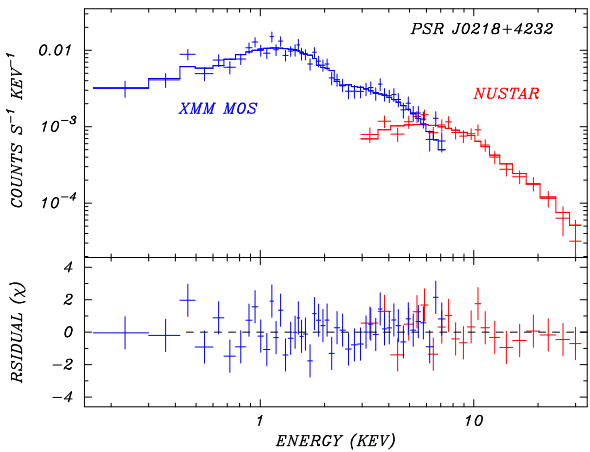}
\hfill
}
\caption{Phase-resolved \nustar\ spectrum of PSR J0218$+$4232 fitted
  simultaneously with the phase-averaged \xmm\ MOS spectrum to an
  absorbed power-law model with independent normalizations.  The upper
  panel presents the \xmm\ ({black}) and \nustar\ ({red})
  spectral data points ({crosses}) along with the best-fit model
  ({solid lines}) given in Table~\ref{tab:spectable}.  The lower
  panel shows the fit residuals in units of sigma.  }
\label{fig:j0218_spec}
\end{figure}

The available archival \xmm\ data of PSR J0218$+$4232 are from a 37~ks
exposure obtained on 2002 February 11 (ObsID 0111100101) and was
originally presented in \citet{Webb04}. The EPIC pn instrument was used in
fast timing mode, while both EPIC MOS cameras were operated in full window
mode. Roughly half of the observation is affected by strong background
flaring, resulting in only 20~ks of usable exposure time. To generate
a photon list suitable for timing and phase-resolved spectroscopy we
followed the same procedure as outlined in \S\ref{sec:method}.

Compared to the X-ray pulse profiles of \psr\ and B1937$+$21, the two
pulse peaks of PSR J0218$+$4232 are appreciably broadened and
their phase separation is the same, to within measurement error.  In
addition, there is evidence for a bridge of emission connecting the
two peaks, with no obvious off-pulse interval.  Unlike the other two
MSPs, the pulsed signal for PSR J0218$+$4232 is not consistently
detectable during a \nustar\ orbit with the same quality, allowing the
possibility of a half-cycle phase ambiguity for some ($\sim$1$-$2)
measurements.  Nevertheless, our approach to correcting the photon
arrival times is able to recover the accurate pulse profile in 0.1~ms
phase bins, as shown in Figure~\ref{fig:j0218_lightcurve}.

\citet{Mineo00} presented evidence for energy dependence of the X-ray
pulse profile of PSR J0218$+$4232 using \textit{ROSAT} HRI and \textit{BeppoSAX} data.
Comparing three energy bands, 0.1$-$2.5 (HRI), 1.6$-$4, and 4$-$10
keV, they report that the measured flux contained in pulse (P2)
increases with energy and is greater than P1 at higher energies.
Furthermore, the D.C. fraction of the bridge emission decreases
linearly with energy.  However, a similar timing study by
\cite{Webb04} using \xmm\ yielded contradictory results, where the P2
is found to be most prominent at the intermediate energy.

In the current study, we cannot address this question definitively
using the \chandra\ and \nustar\ data as it is not possible to phase align
these data in an absolute sense.  However, within the limits of photon
statistics, the background-subtracted HRC pulse profile, whose counts
are mostly $<2$~keV, is statistically identical to the background-subtracted 
\nustar\ ($>3$~keV) pulse profile when aligned on the larger
peaks (as shown in Figure \ref{fig:j0218_lightcurve}). Furthermore, if we
compare the (manually aligned) \nustar\ data in four energy bands, $<$4,
4$-$10, 10$-$20, and 20$-$30~keV, we find no significant evolution of
the pulse profile or D.C. component with energy. We conclude that
there is no convincing evidence for an energy-dependent pulse profile
to date.


For a spectral analysis of PSR J0218$+$4232 we again extracted a
\nustar\ spectrum for the pulsed emission only, centered on the two
pulse peaks ($3+4$ of 20 phase bins). Due to a lack of a clear off-pulse
interval to represent the background, we used the phase-averaged
background spectrum, scaled to the on-pulse interval.  We fit this
\nustar\ spectrum simultaneously with the phase-averaged \xmm\ EPIC
MOS spectra and, as expected, an absorbed power-law model results in a
good fit with $\Gamma=1.10\pm0.09$, $N_{\rm H}=(5.2\pm3)
\times10^{20}$ cm$^{-2}$, with $\chi_{\nu}^2 = 0.88$ for 63~dof (see Figure 9). By
comparing the \xmm\ and \nustar\ best-fit fluxes, we find that the
bridge emission contributes $\lesssim35\%$ to the total in the
2$-$10~keV band. The lower limit allows for possible imperfections in
the \nustar\ timing reconstruction, which might add to the bridge
emission. Our spectral results are consistent with those reported in
\cite{Webb04}.

\begin{deluxetable*}{lcccccl}
\tablecolumns{7} 
\tablewidth{0pt}  
\tablecaption{Summary of phase-resolved spectroscopic analysis for the three MSPs\tablenotemark{a}\label{tab:spec}}  
\tablehead{\colhead{Pulsar} & \colhead{$N_{\rm H}$} & \colhead{$\Gamma$} & \colhead{$F_X$\tablenotemark{b}} & \colhead{$F_X$\tablenotemark{b}} & \colhead{$F_X$\tablenotemark{b}} & \colhead{$\chi^2/$dof}  \\
           \colhead{}       & \colhead{($10^{21}$ cm$^{-2}$)} & \colhead{}         &  \colhead{(0.3--8 keV)} &  \colhead{(2--10 keV)}   & \colhead{(3--79 keV)} }
cp \startdata
B1821$-$24 \chandra +\nustar\        & $4.01\pm0.04$  & $1.28\pm0.05$  & $3.19\pm0.09$[C] & $3.69\pm0.19$[N] & $22.1\pm1.5$[N] & $0.91/226$ \\  
B1937$+$21 \chandra +\xmm +\nustar\ & $17.8\pm2.7$  & $1.16\pm0.11$  & $3.01\pm0.18$[X] & $2.87\pm0.21$[X] & $18.2\pm2.8$[N] & $0.91/83$ \\  
J0218$+$4232 \xmm +\nustar\           & $0.52\pm0.3$ & $1.10\pm0.09$  & $3.97\pm0.26$[X] & $3.88\pm0.35$[X] & $19.8\pm2.3$[N] & $0.88/63$ 
\enddata
{\bf Notes.}
\tablenotetext{a}{Simultaneous fits for the combined data sets with independent normalizations. Fits to individual data sets are presented in the text.}
\tablenotetext{b}{Unabsorbed flux in units of $10^{-13}$ erg cm$^{-2}$ s$^{-1}$, for the [C] \chandra, [X] \xmm, or [N] \nustar\ spectral component of the composite fit.}
\label{tab:spectable}
\end{deluxetable*}

\section{MSP  Spectral Energy Distribution}

Using the improved constraints on the hard portion of the X-ray
spectra for the three MSPs, we can re-examine the SEDs of these objects 
ranging from soft X-rays to GeV
gamma-rays by combining our X-ray results with archival spectra in the
0.1$-$100~GeV band obtained with the \textit{Fermi}-LAT. Generally, the
$\gamma$-ray spectra of MSPs are well fit by a power-law model with an
exponential cutoff, most typical of  rotation-powered pulsars.

\citet{Johnson13} reported on the detection of $\gamma$-ray pulsations
from \psr\ at a level of 5.4 $\sigma$ in \fermi\-LAT data.  These
pulsations were misaligned from the radio and X-ray peaks, as is
common among MSPs. Its $\gamma$-ray spectrum is well fit by a
power-law with index $\Gamma=1.6\pm0.3$ and cutoff energy $E_c =
3.3\pm1.5$ GeV. Similarly for PSR B1937$+$21, an analysis of
\textit{Fermi}-LAT data by \citet{Guillemot12} find pulsed emission
with power-law index $\Gamma=1.43\pm0.87$ and a fairly low cutoff
energy of $E_c = 1.15\pm0.74$ GeV.  PSR J0218$+$4232, one of the
first MSPs detected with \textit{Fermi} LAT, is characterized by a
power-law index $\Gamma=2.0\pm0.1$ and cutoff energy $4.6\pm1.2$ GeV
\citep{Abdo13}.

\begin{figure*}[t!]
\begin{center}
\includegraphics[angle=0,width=0.75\textwidth]{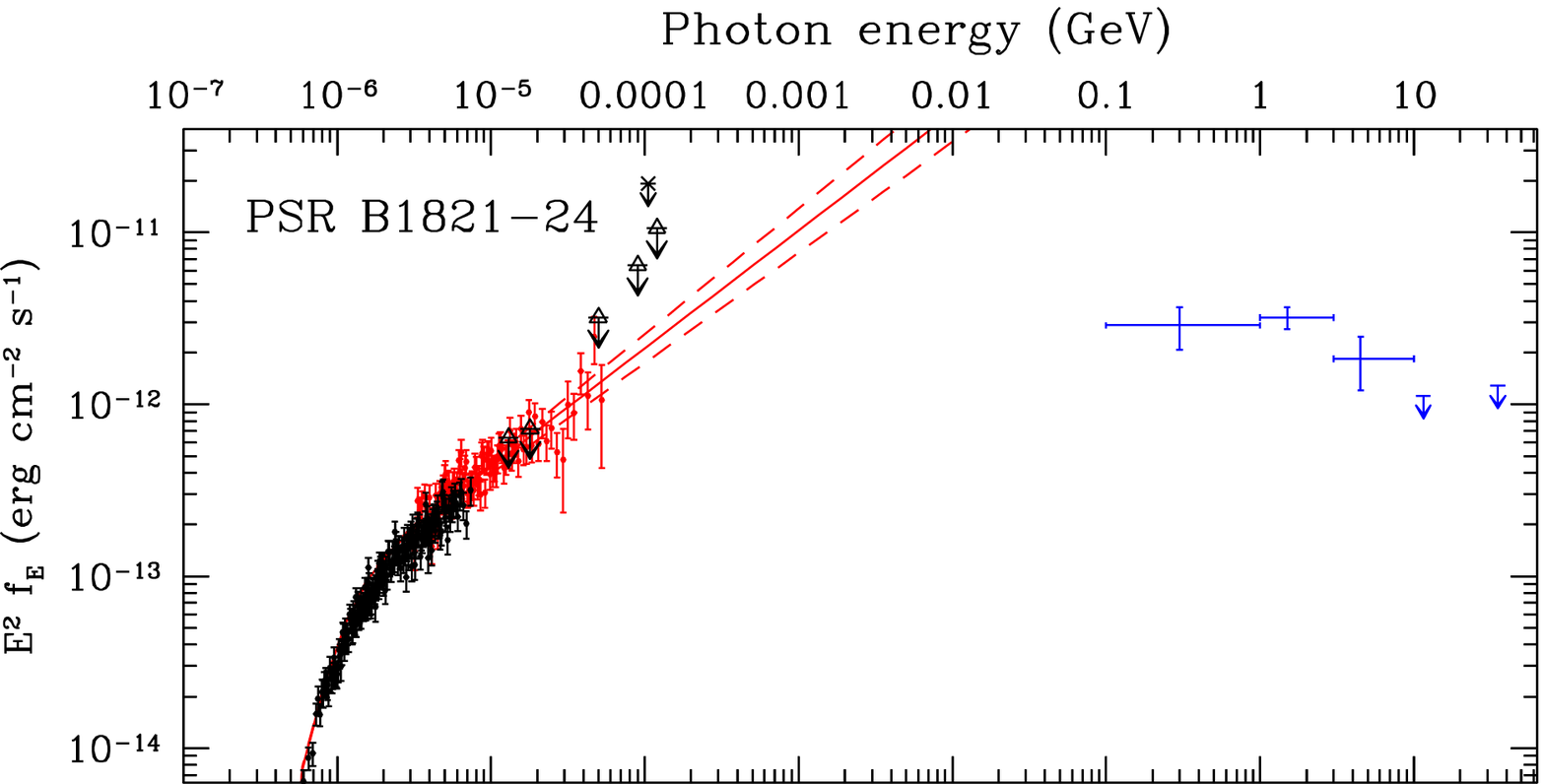}
\includegraphics[angle=0,width=0.75\textwidth]{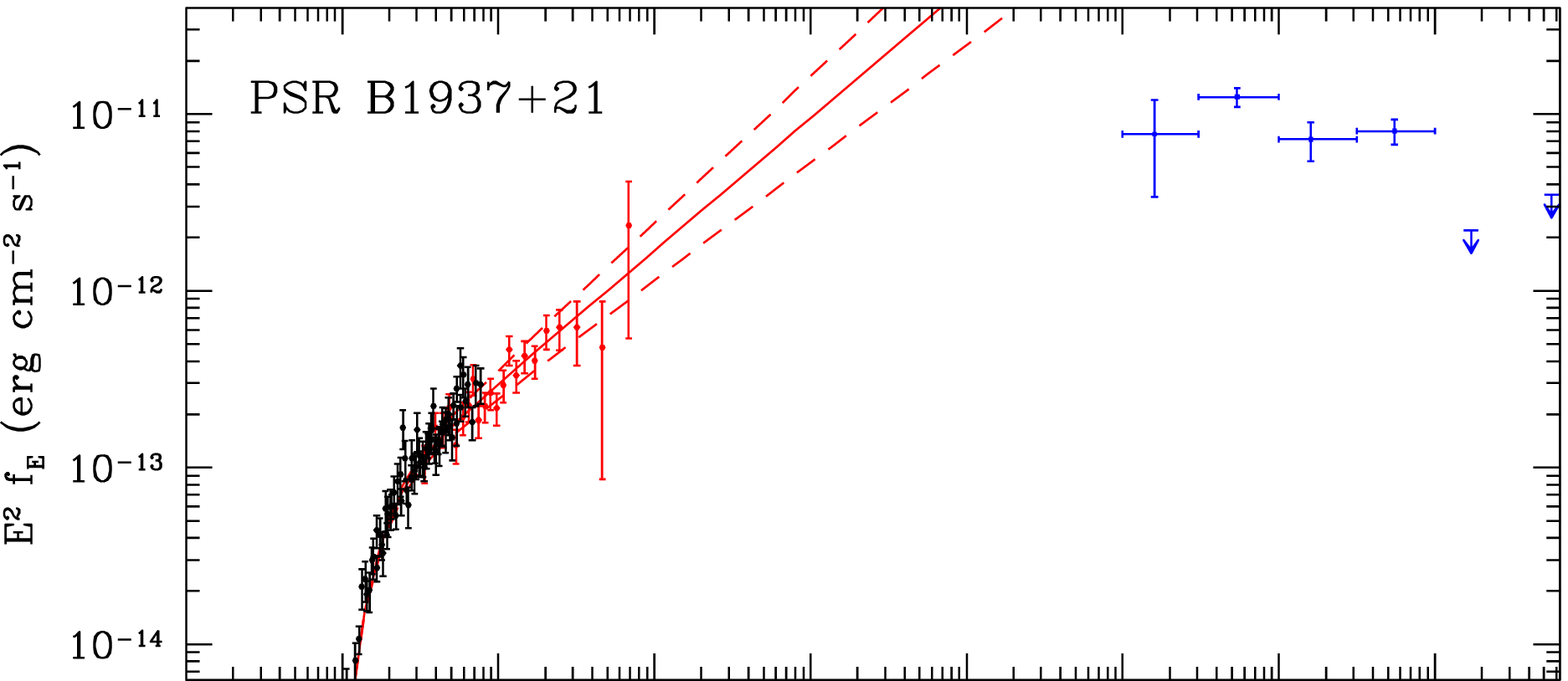}
\includegraphics[angle=0,width=0.75\textwidth]{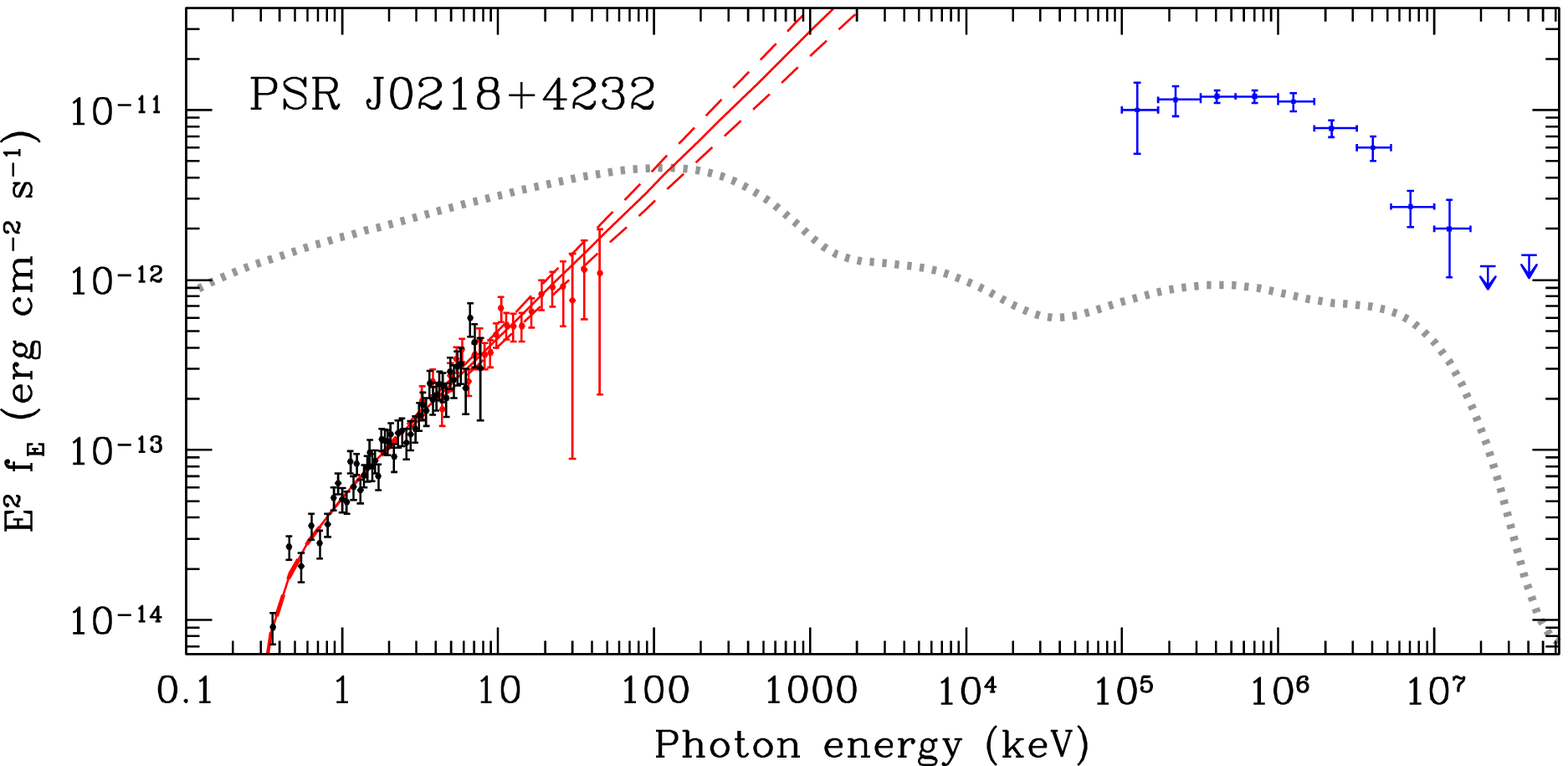}
\caption{The spectral energy distribution of the three MSPs,
  spanning the soft X-ray to high-energy $\gamma$-ray range. The {
    solid lines} represent the respective unfolded spectra from the
  fits presented herein. The {dashed lines} are the extrapolated
  90\% C.L. uncertainties.  The {dotted line} at low energies
  shows the unabsorbed extrapolation. The {blue crosses} and {
    upper limit arrows} show the \textit{Fermi}-LAT data for photon
  energies $\ge$100 MeV. The open triangles and cross show the
  \textit{HEXTE} and \textit{OSSE} upper limits from \citet{Kuiper04}.
  {top} to {bottom} panels: \psr, PSR B1937$+$21, and
  PSR J0218$+$4232.  The \textit{Fermi}-LAT spectra are from
  \citet{Johnson13}, \citet{Ng14}, and \citet{Abdo13}, respectively. The
  Crab pulsar SED ({\it grey line}) with an arbitrarily rescaled flux is shown for comparison in the
  {bottom} panel.}
\label{fig:bband}
\end{center}
\end{figure*}

The SED of the three MSPs spanning 0.1 keV to $\sim$60 GeV are shown
in Figure~\ref{fig:bband}. It is apparent that extrapolation of the
power-law spectrum from the X-ray range to higher energies greatly
overpredicts the flux in the \textit{Fermi}-LAT band at 100 MeV and above 
for all three MSPs. The extrapolated X-ray spectrum is already larger than the flux in the lowest \textit{Fermi}-LAT energy bin starting at $\approx$100 keV, $\approx$1 MeV, and $\approx$400 keV for the three MSPs, respectively. This clearly indicates that a break or smooth turnover 
must occur between the \nustar\ and the \textit{Fermi} bands.  The addition of the \nustar\
data, whose spectrum shows no conclusive evidence for deviation from a
simple power law, requires that any spectral turnover must occur
above $\sim$100 keV. Due to the absence of a telescope with sufficient
sensitivity between 100~keV to 100~MeV, we cannot constrain the
spectrum of the three MSPs in this range. For \psr, the presently
available data with energies $\gtrsim$100 keV from \textit{HEXTE} and \textit{OSSE} \cite[from][]{Kuiper04} only provide upper limits on the
flux that are not in conflict with the extrapolation of the X-ray
power law.

The  spectral energy distributions of the three MSPs are
reminiscent of the high-energy emission from the Crab pulsar. In the
$\sim1$$-$50~keV band, the Crab pulsar also shows a purely non-thermal
spectrum with a power-law slope of $\Gamma \sim 2.1$, but evidence of a
slight curvature. If curvature is present in the intrinsic spectrum
of our MSPs, it is not discernible at a significant level with the
current \nustar\ data.  Above $\sim$100~keV, the Crab pulsar spectrum
turns over multiple times before increasing in the \fermi\ band
\citep[see bottom panel of Figure~\ref{fig:bband}, as well as ][and references therein]{Buhler14}. 
It is likely that the three MSPs exhibit similar spectral behavior in the 100 keV to 100 MeV range.

There exist a variety of theoretical models of pulsar electrodynamics
that are used to interpret the observed high-energy magnetospheric
emission from pulsars.  They differ principally in the assumed
location of particle acceleration and attendant emission: polar cap
\citep[e.g.,][]{Daugherty96}, outer gap \citep[e.g,][]{Romani96}, slot
gap \citep[][]{Muslimov04}, or beyond the light cylinder in the
current sheet \citep[see, e.g.,][]{Cerutti16}.  In the context of
these models where the pulsed high-energy emission originates within
the magnetosphere, the observed SEDs can be understood in terms of a spectrum
transitioning from a synchrotron component at soft and hard X-ray
energies to one dominanted in $\gamma$-rays by either curvature radiation or
synchrotron emission from the current sheet.  The improved constraints
on the hard X-ray spectra of the three MSPs considered here provide
additional information regarding the physics of the pulsar
magnetosphere. Any viable model needs to account for both the pulse
shapes and the spectral energy distribution we have reported here.

\section{CONCLUSIONS}

We have presented detailed broadband X-ray timing and spectroscopic
analyses of the magnetospheric-dominated MSPs, PSRs B1821$-$24,
B1937$+$21, and J0218$+$4232. By making use of the sharpness of their
pulses, bright enough to detect in individual telescope orbits, we are able to
phase align the pulse signal between orbit gaps. By this means we are
able to negate the significant drift in the \nustar\ clock during these
gaps, likely caused by solar illumination. This allowed a
phase-resolved analysis of all three pulsars.

It is important to note the limitations of this approach. The applied
clock corrections are only valid during the course of an observation
since the clock drift by $\sim$milliseconds occurs on timescales
comparable to the \nustar\ orbit ($97$ minutes). In addition, the
clock-correction method benefits from sharp pulsations; for fainter
pulsars with broader modulations, the pulsations may not be detectable
on short (orbit) intervals (see, e.g., \citealt{Guillot16} for the
case of PSR J0437$-$4715). Imaging and temporal analysis show that
the X-ray signals for \psr\ and B1937$+$21 are 100\% pulsed, and there 
is no evidence for energy dependence in this energy band.

Spectra obtained with \nustar\ allow us to measure all three pulsars
to higher energies than possible with \xte, in comparable exposures,
highlighting the advantage of focusing optics, which results in
a substantially lower background.  At these higher energies we find the
same power law seen at lower energies with no conclusive evidence for
a spectral break or turnover.  Furthermore, the pulse profiles for all
three pulsars are found to be essentially invariant with energy, from
the soft to the hard X-ray bands.  Combined with the GeV data from
\textit{Fermi}-LAT, we have obtained the best  measurements yet of the
broadband spectral shape and high-energy pulsed emission of each
MSP. We find that in all cases the power-law spectrum needs to turn
over somewhere above the \nustar\ band and below 100~MeV to be
consistent with the high-energy $\gamma$-ray spectrum.  This can be
understood in terms of a spectrum transitioning from a dominant
synchrotron component in X-rays to curvature radiation or current
sheet emission in $\gamma$-rays, very similar to the Crab pulsar.

\acknowledgements We greatly appreciate discussions with M.~Bachetti
and C.~Markwardt concerning \nustar\ clock timing issues, and K.
Forster for the necessary assistance in planning the \psr\ observations. 
We thank A.~Harding for discussions regarding high-energy magnetospheric emission from pulsars.
We also thank S.~Ransom for supplying the radio ephemeris of \psr.  This
work was supported under \textit{NuSTAR} Cycle 1 Guest Observer
Program grant NNX15AV29G awarded through Columbia University.  E.V.G.
thanks Josep Maria Paredes for hosting his sabbatical at the
University of Barcelona Institut de Ci\`encies del Cosmos (ICCUB) and
acknowledges support through the ''Programa Estatal de Foment de la
Investigaci\`o Cient\'{\i}fica i T\`ecnica d'Excell\`encia,
Convocat\`oria 2014, Unitats d'Excell\`encia {\it Maria de Maeztu}."
The \nustar\ mission is a project led by the California Institute of
Technology, managed by the Jet Propulsion Laboratory, and funded by
the National Aeronautics and Space Administration. We thank the
\nustar\ Operations, Software, and Calibration teams for support with
the execution and analysis of these observations. This research made
use of the \nustar\ Data Analysis Software (NuSTARDAS) jointly
developed by the ASI Science Data Center (ASDC, Italy) and the
California Institute of Technology (USA).  The scientific results
reported in this article are based in part on data obtained from the
\chandra\ Data Archive. A portion of the results presented are based
on observations obtained with \textit{XMM-Newton}, an ESA science
mission with instruments and contributions directly funded by ESA
member states and NASA.  This research has made use of data and
software provided by the High Energy Astrophysics Science Archive
Research Center (HEASARC), which is a service of the Astrophysics
Science Division at NASA/GSFC and the High Energy Astrophysics
Division of the Smithsonian Astrophysical Observatory.  We also
acknowledge extensive use of the arXiv and the NASA Astrophysics Data
Service (ADS).

Facilities: \textit{NuSTAR, CXO, XMM.}

\end{document}